\documentclass[submission,copyright,creativecommons]{eptcs}
\providecommand{\event}{FMAS 2022} 
\usepackage{setspace} 
\usepackage{graphicx}
\usepackage{subcaption}
\usepackage{amssymb}
\usepackage{amsmath}
\usepackage{xspace}
\usepackage{csquotes}
\usepackage{caption}
\usepackage{subcaption}
\usepackage{sty/shortcuts}
\usepackage{hyperref}
\usepackage{siunitx}
\def\titlerunning{A Doxastic Characterisation of Autonomous Decisive Systems}
\def\authorrunning{A. Rakow}
\begin{document}
\title{A Doxastic Characterisation of Autonomous Decisive Systems\thanks{
	This work is partly funded by the German Research Council (DFG) as part of
	the project “Integrated Socio-technical Models for Conflict Resolution 
	and Causal
	Reasoning”, grant no. FR 2715/5-1, DA 206/12-1}}
%
%
\author{Astrid Rakow
%
%
\institute{Carl von Ossietzky University, 26129 Oldenburg, Germany}
\email{a.rakow@uol.de}}
\maketitle              
\begin{abstract}
	A highly autonomous system (\HAS) has to assess the situation it is in and derive
 beliefs, based on which, it decides what to do next. 
	The beliefs are not solely based on the observations the \HAS has made so far, but also on general insights about the world, in which the \HAS operates. These insights have either been built in the \HAS during design or are provided by trusted sources during its mission.
	Although its beliefs may be imprecise and might bear flaws, the \HAS will have to extrapolate the possible futures in order to evaluate the consequences of its actions and then take its decisions autonomously.
	In this paper, we formalize an \emph{autonomous decisive system} as a system that always chooses actions that it currently believes are the best. 
	We show that it can be checked whether an autonomous decisive system can be built given an application domain, the dynamically changing knowledge base and a list of LTL mission goals. 
	We moreover can synthesize a belief formation for an autonomous decisive system.
	For the formal characterization, we use a doxastic framework for safety-critical \HASes where the belief formation supports the \HAS's extrapolation.
\end{abstract}

\section{Introduction}
\normalsize
Nowadays highly autonomous systems (\HASes) are already able to work within complex environments as e.g. the autonomous Tesla cars and the NASA's helicopter \enquote{Ingenuity} demonstrate. 
On the other hand, these systems have to satisfy high safety standards. This is especially so for human cyberphysical systems, where humans and autonomous systems symbiotically cooperate \cite{HCPS}.
Accidents like \cite{Uber18,accidents2020} show that there is still a long way to go before the high safety standards can be reliably implemented.%

A \HAS faces the great challenge of making good decisions despite an overwhelmingly complex and dynamic environment.
In order to decide what to do next, a \HAS builds a belief about its current situation, usually by combining incoming sensor data with prior data.
As this data is often incomplete and data from different sources may be contradictory, the \HAS can never be sure of its current state and thus deems several variants of the situation possible.
The \HAS extrapolates the possible futures that might develop from the current situation. 
Therefore, it simulates its action alternatives on a world model that describes the impact of its actions as well as how the environment might evolve.
Hence, a \HAS has to take decisions based on beliefs that concern its past, present and future  and that can only approximate the real world.

In this paper, we focus on safety critical systems and formalize the notion of autonomous-decisive system.  
We call a system autonomous-decisive, if it can choose its actions based on the contents of its beliefs and then achieves its goals within the intended application domain.
We formalize the notion within a game-theoretic doxastic framework, where
we use the possible worlds semantics~\cite{reasoningA} to represent the beliefs of a system.
A so-called \emph{belief formation function} specifies the belief for a given history of perceptions 
 and the current knowledge base.
The latter holds knowledge and hard beliefs. 
It captures constraints that an engineer chooses to enforce on the system's beliefs (e.g. insights about the application domain) as well as statements that the \HAS gets from trusted sources during its mission. 
As we focus on safety-critical systems, we use a two-player zero-sum reactive game. 
That is, we consider the environment 
 as adversarial and the system has to operate successfully irrespective of the environment's choices.
By considering the belief formation itself as a special kind of observation-based strategy, we can synthesize a witness belief formation that enables a \HAS to autonomously achieve its goals.

Conceptually, the presented formalization will be useful in the early design, where simple and abstract models are considered and decisions about the resources of a system like its sensors and capabilities are made.
The formalism is applicable to any system that has to take autonomous decisions. 
We imagine that the application domain and the mission goals are formally described. 
When during the design the possible beliefs of the system have been fixed, 
our framework can be used to examine whether the system can successfully take autonomous decisions within the targeted application domain.  
When the setting does not allow to decide autonomously, the design of its possible beliefs, the system's sensing capabilities or actions might have to be adjusted. In certain cases, even the model of the application domain has to be revised~\cite{DF11}.

\begin{sloppypar}
\paragraph{Outline}Next we introduce the ingredients of our formal framework.
In \autoref{sec:autonom} we formalize autonomous-decisive systems as systems that are able to take decisions based on their internal assessment of the situation. 
We prove that it is decidable whether such a system exists for a given application domain, available knowledge/hard beliefs, possible beliefs and mission goals. 
In \autoref{sec:relwork} we present related and future work and conclude in \autoref{sec:concl}.
\end{sloppypar}

\section{Ingredients of our Doxastic Framework}\label{sec:ingredients}
\normalsize
In this section, we will introduce the ingredients our framework alongside a running example.
As a toy example, we consider the set-up as sketched in \autoref{fig:setup}(a) where two cars are on separate lanes heading towards each other. 
The left car is called \ego and is our \HAS. It has to avoid collisions and it has to take the left turn.
The right car, \other, is uncontrolled (from \ego's perspective).

\myparagraph{A World}
We assume that 
the \HAS's intended application domain is captured via a Kripke structure, modelling static and dynamic objects and their interactions.
This so called \emph{design-time world} \universeD captures the information about the application domain as well as test criteria that the \HAS has to master as part of the development and certification process. 

\begin{definition}{world}
	Formally, a \emph{world} \universe is a labelled Kripke structure 
	$\universe=(\States,\Edges,\Label,\Init)$, 
	where \States is the set of states, 
	$\Edges\subseteq\States\times\States$ are edges between states, 
	$\Label=\Label_\States\cup\Label_{\Edges}$ where
	$\Label_\Edges=\Edges\rightarrow {2^{\Act}}$ labels edges with  a subset of the finite set of actions $\Act$ and $\Label_\States:\States\rightarrow {2^{\Props}}$ labels states with valuations of a finite set of Boolean propositions \Props, and $\Init\subseteq\States$ is the set of initial states. 
	$\Act=\Act_{\ego}\times\Act_{\env}$ is a set of tuples defining the simultaneous actions of our autonomous agent \ego and its environment \env, which may include other agents.
	We assume that the transition relation is defined for all states and actions, i.e., $\forall \state\in\States,\forall \act\in\Act,\exists \edge\in\Edges: \act\in\Label_\Edges(\edge)$.
\end{definition}
In order to express that an action is not enabled at \state, \universe can transition into a dedicated state $\sundef$ that is accordingly labelled.\footnote{In this paper any strategy has hence to avoid $\sundef$.}
A sequence of states $\Path=\state_0\state_1\ldots\state_n\in\States^{*}\cup\States^{\omega}$ is a \emph{path} in \universe iff $\forall i, 0\leq i<|\Path|: (\state_{i},\state_{i+1})\in\Edges$. $\Path(i)$ is the state $\state_i$, $\Path_{\leq m}$ denotes the prefix $\state_0\ldots\state_m$ and $\last(\Path)$ is the last state of a finite path $\Path$. $\Path$ is initial iff $s_0\in\Init$.

%
\begin{example}{A world}\label{ex:uni}
	In our running example, the actions of \ego are \textsf{f}, \enquote{moving one step forward if possible}, \textsf{t}, \enquote{turn and move one step forward}. 
	\Other is either a slow car and moves one tile forward, if possible, or  it is a fast car.  
	In that case, it moves two positions forward, if it is at its initial position, else it moves one tile forward, if possible. \Other's actions are \textsf{f} and \textsf{F}, \enquote{move two positions forward}. The actions of \ego and \other are annotated by pale blue and dark blue arrows in \autoref{fig:setup}(a).

	The propositions $\Props_{\textit{pos}}=\Props_{\xego}\cup\Props_{\yego}\cup\Props_{\xother}$ with  $\Props_{\xego}=\{\xego=i|1\leq i\leq 4\}$, $\Props_{\yego}=\{\yego=i|1\leq i \leq 3\}$, $\Props_{\xother}= \{\xother=i|1\leq i \leq 4\}$ encode the positions of the two cars, where $\xego=i$ and $\yego=i$ represent the horizontal and vertical position of \ego, and $\xother=i$ represents the horizontal position of \other. Its vertical position is always two. 
	\Other's car type is encoded via the propositions \textsf{slow} (s) and \textsf{fast} (f). 
	We assume that \ego cannot observe \other's car type directly, but it has sensors perceiving \other's colour, which is either red or blue. 
	The proposition \blue (\red, \textsf{fast}, \textsf{slow}) is true, iff \other is a blue (red, fast, slow) car. 
	The propositions \bp and \rp encode what \ego perceives as \other's colour. 
	They are used to modeling \ego's imperfect colour recognition,  while the propositions \blue and \red encode the true colour of \other. 
	We assume that \ego's colour perception works correctly, when \ego and \other are less than two tiles apart, otherwise the sensor switches colours ($\bp=\neg\blue$, $\rp=\neg\red$). Let $\Props_{\textit{cartype}}$ be the set $\{\textsf{fast, slow, blue, red, blue$_p$, red$_p$}\}$. 
	The propositions in our example are hence $\Props=\Props_{\textit{pos}}\cup\Props_{\textit{cartype}}\cup\{\textsf{undef}\}$, where \textsf{undef} labels the sink state.

	\begin{figure} 
		\addtocounter{figure}{1}    
		\begin{minipage}{0.4\textwidth}
			\vspace*{10mm} 
			\centering
			\includegraphics[width=\textwidth]{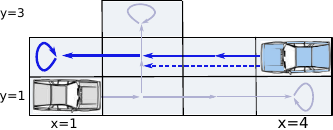}\\
			{(a) Sketch of the world}\label{fig:sketch}
		\end{minipage}
		\hfill
		\begin{minipage}{0.45\textwidth}
			\centering
			\includegraphics[width=\textwidth]{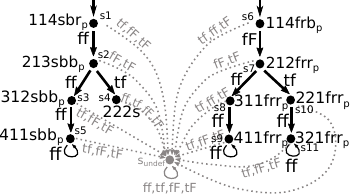}\\
			{(b) World model}\label{fig:wm}
		\end{minipage}
		\begin{minipage}{0.3\textwidth}
			\footnotesize
		\end{minipage}
		\addtocounter{figure}{-1}    
		\caption{A simple world  \universeD.}	
		\label{fig:setup}
		\label{fig:wm}
	\end{figure}
	Figure \ref{fig:setup}(b) shows the Kripke structure of our design time world.
	States are labelled with the propositions that hold in the respective state. 
	The label $\mathsf{abcdef}\in\mathbb{N}^3\times\{f,s\}\times\{b,r\}\times\{b_p,r_p\}$  encodes that $\xego=a,\yego=b,\xother=c$ are true and \other's car type is \textsf{d}, its  colour is \textsf{e} whereas the perceived colour is \textsf{f}. 
	The valuations of all other propositions are false. 
	Likewise, the label \textsf{undef} encodes that the only valid proposition is \textsf{undef}.   
	Edges are labelled with sets of actions. We omit the sets' brackets.
	For example, the label \textsf{ff} denotes the set $\{\textsf{ff}\}$, that contains the one action \textsf{ff}, where \ego and \other simultaneously move one step forward, if possible.   
	Actions that are not enabled at a state lead to the sink state \sundef. 
	We omit the sink state and sink transitions in the sections that follow.
\end{example}

%
%
%
\myparagraph{Goal List}
Our \HAS, \ego, has to achieve a prioritized list of goals during a manoeuvre.
A \emph{goal} \goal is a linear-time temporal logic (LTL) formula~\cite{PrinciplesOfMC}. 
We denote the temporal operators \enquote{globally} by $\Box$, \enquote{eventually} by $\Diamond$, \enquote{next} by \Next and \enquote{until} by \Until. 
We interpret the LTL formulae over (infinite) traces, which are infinite
sequences $\cmp = \cmp_0 \cmp_1\ldots \in (2^\Props)^\omega$ of valuations of \Props. 
Satisfaction of an LTL formula $\varphi$ by
a trace $\cmp$ is denoted as $\cmp \models \varphi$.
A \emph{goal list} $\goalList = (\Goals, \prio)$ consists of a set $\Goals$ of LTL formulae and a priority function $\prio : \Goals \rightarrow \{1, \ldots, |\Goals|\}$ where
$\goal\in\Goals$ is more important than $\goal'\in\Goals$ iff $\prio(\goal)<\prio(\goal')$.
We say that a trace $\cmp$ satisfies \goalList with priority up to $n$ if $n$ is the highest priority up to which \cmp satisfies all goals of \goalList, i.e., $\cmp \models \goal$ for all $\goal \in \Goals$ with $\prio(\goal) \leq n$.  
A set of traces \Cmp satisfies \goalList up to $n$, if all $\cmp\in\Cmp$ satisfy \goalList up to a priority $m$ with $n\leq m$ and $n$ is the greatest such priority. For convenience we extend goal lists to include the trivial goal \true, i.e., $\true\in\Goals$ and $\prio(\true)=0$.

\begin{example}{Prioritized Goals}\label{ex:goals}
	We formalize collision freedom as $\varphi_{c}=\Box (\xego=2\land \yego=2\Rightarrow \xother\not=2)$, and $\varphi_{t}=\Diamond (\yego=3)$ expresses that \ego eventually does the turn. 
	In order to rule out that disabled transitions are taken, we use the goal $\varphi_{u}=\Box \neg \textsf{undef}$.
	The priorities are given by $\prio(\varphi_u)=1,  \prio(\varphi_c)=2, \prio(\varphi_t)=3$.

	Let us now take a closer look at what \ego should do in order to accomplish its goals.
	If \other is slow, then \ego should not take the turn, but instead it should drive straight on, in order to avoid the collision. 
	If \other is fast, then  \ego can take the turn and accomplish all its goals.
\end{example}

\myparagraph{Observations}
\Ego, being highly autonomous, will take decisions based on the beliefs that it has constructed about the world in which it operates. 
\Ego derives its beliefs from the observations made so far and the knowledge/hard beliefs it has about the world. 

A world is usually  only partially perceivable by \ego via observations. \emph{Observations} \Obs are propositions of \universeD whose valuations \ego can assess and that represent  e.g.  sensor readings or received messages from other agents. Observations shed light on \universeD, but they do not have to be truthful, as illustrated in \autoref{ex:uni}, where initially \bp,\rp were switched and hence \ego does not perceive the correct colour all the time. 
Nevertheless, a \HAS has to reason about the current state of the world based on its past observations.
Let $P$ be a set of propositions, $P\subseteq  \Props$. 
We call $h$ a ($P$-)\emph{history} of a state \state, if there is an initial path $\Path$ of \universe leading to \state and
$h=h_0h_1\ldots h_n$ is the sequence of state labels along $\Path$, $h_i=\Label(\Path(i))\cap P, \forall i, 0\leq i\leq n$. 
We denote the set of $P$-histories as $\histories_{P}$.  
We say $h$ is \emph{observable} iff $P\subseteq \Obs$.
\begin{example}{Observable History}\label{ex:hist}
	In our example \ego cannot observe \other's position due to a broken distance sensor, but it can observe its own position and the colour of \other, so that $\Obs:=\Props_{x_e}\cup\Props_{y_e}\cup\{\textsf{undef, \bp, \rp}\}$.
	Given the world of \autoref{fig:setup}(b), \Small{$h\,=\,$114sbr$_p,$213sbb$_p,$312sbb$_p$,411sbb$_p$} is the history along the path \Small{s1$,$s2$,$s3$,$s5} wrt \Props, whereas \Small{11r$_p$,21b$_p$,31b$_p$,41b$_p$} is the observable history wrt \Obs.
\end{example} 
%

\myparagraph{Beliefs} A belief describes what \ego currently thinks. 
For instance, \ego may think that it saw an approaching vehicle and  that this vehicle is a slow car (or a fast car). 
Depending on the type of car, \ego will imagine different possible future evolutions of the current situation. 
\begin{definition}{belief, reality}
	A \emph{belief} \belief is the set of alternative realities that \ego currently deems possible, $\belief=\{\reality_0,\ldots,\reality_n\}$. 
	A \emph{reality} is a pair $\reality=(\world,\cStates)$ of a \emph{(possible) world} \world and a set of believed current state $\cStates\subseteq\States$, where any current state is reachable from  one initial state but no initial state is reachable from a current state.
\end{definition}
A reality describes possible pasts, current states and futures: pasts are captured by the set of paths between initial states \Init and current states \cStates, the possible futures are paths from the current states. 
\begin{example}{Alternative Realities and Beliefs}\label{ex:altBel}
	To illustrate the notion of belief as introduced above, let us consider two alternative realities of \ego in \autoref{fig:ex_bel}(a)+(b). The believed past is marked by framing state labels.\\[2mm] 
	\begin{minipage}{\textwidth}
		\begin{minipage}{.59\textwidth}
			\begin{minipage}{0.48\textwidth}
				\addtocounter{figure}{1}    
				\centering		
				\includegraphics[width=.78\textwidth]{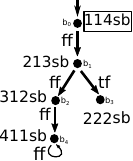}\\
				{(a) \Small{\ego is at $b_0$, \other is slow}}
			\end{minipage}
			\hfill
			\begin{minipage}{0.48\textwidth}
				\centering
				\includegraphics[width=0.78\textwidth]{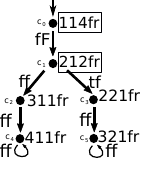}\\
				{(b) \Small{\ego is at $c_1$, \other is fast.}}
			\end{minipage}
			\addtocounter{figure}{-1}    
			\hspace*{-2mm}
			\captionof{figure}{Two alternative realities of \ego\vspace*{3mm}}\label{fig:ex_bel}
		\end{minipage}
		\hfill
		\begin{minipage}{0.38\textwidth}
			{
				The alternative reality of (a) describes that \other is slow and that \ego itself is at the initial state $b_0$ of that world.
				In (b) \other is fast and \ego is at $c_1$, the \enquote{second} state of that world.
				The believed history is \Small{\textsf{114fr},\textsf{212fr}}.
				}
				{
					Since a belief is a set of alternative realities, singletons of either (a) or (b) form a belief. Also, the set of (a) and (b) forms a belief, where \ego thinks both alternatives are possible.
					}

		\end{minipage}
	\end{minipage}
	\null\enspace
\end{example}
%
Note, \ego may believe in possible worlds that are substantially different from  \universeD, since it usually captures the real world by simplified concepts and rules, that may reflect the inner workings of the application domain only coarsely but sufficiently. 
Since a \HAS has only finite resources, we assume that the set of \emph{possible beliefs} $\Beliefs$ and the set of \emph{possible worlds} $\Worlds:=\bigcup_{(\cStates,\world)\in \Beliefs}\world$ are finite. 
The choice of \Beliefs constitutes a design decision within the development process of the \HAS.
\begin{example}{Possible Beliefs \Beliefs}\label{ex:bel}
	Our \ego has been designed to represent a certain set of scenarios, for which it can evaluate what to do by extrapolating the future. 
	Figures \ref{fig:bel_setup}(a)-(d) sketch a set of the alternative realities via their initial set-up. 
	The other car may be fast or slow, the road may be up to 6 tiles long, the intersection may be at \Small{$x=2$} or \Small{$x=3$}, and the start position of \other varies from \Small{$x=4$} to \Small{$x=6$}.
	Note, that \universeD of \autoref{fig:wm} is described by \autoref{fig:bel_setup}~(a). 
	\begin{figure}[htbp]
		\addtocounter{figure}{1}    
		\begin{minipage}{0.34\textwidth}
			\centering
			\includegraphics[width=\textwidth]{pics/bel_setupII.pdf}\\
			(a)
		\end{minipage}
		\hfill
		\begin{minipage}{0.38\textwidth}
			\centering
			\hspace*{-13mm}\includegraphics[width=\textwidth]{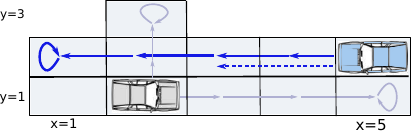}\\
			(b)
		\end{minipage}\\[3mm]
		\begin{minipage}{0.34\textwidth}
			\centering
			\includegraphics[width=\textwidth]{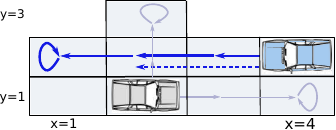}\\
			(c)
		\end{minipage}	
		\hfill
		\begin{minipage}{0.45\textwidth}
			\centering
			\includegraphics[width=\textwidth]{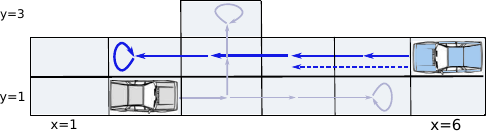}\\
			(d)
		\end{minipage}
		\addtocounter{figure}{-1}    
		\caption{Sketch of \ego's possible beliefs. If \other is fast, it uses the dashed arrow at first and then the solid arrows. If it is slow, it uses the solid arrow.} \label{fig:bel_setup}
	\end{figure}	
	Let \ego's possible beliefs \Beliefs be the beliefs that canonically evolve from these initial scenarios. 
\end{example}

\myparagraph{Knowledge/Hard Beliefs} 
Knowledge/Hard Beliefs are statements with which an engineer equips a system during the system design or that are provided by certain trusted sources (e.g. a traffic control system) during the \HAS's mission.
We think of statements like $(\varphi_a)$ \emph{\enquote{In settled areas the speed limit is \SI{50}{km/h}}}, $(\varphi_b)$ \emph{\enquote{I will be on the highway for the next 20 mins.}} or rules like $(\varphi_c)$ \emph{\enquote{If A promises to give way, I can rely on it.}}. 
We specify these statements via an LTL variant, which we call Belief-LTL~(BLTL).
A BLTL formula $\varphi$ is an LTL formula that is preceded by either the operator $\K$ or $\B$. 
$\K\varphi$ reads as \emph{\enquote{\ego believes that it knows $\varphi$}}. 
A belief \belief satisfies $\K\varphi$, $\belief\models\K\varphi$, if in all alternative realities of \belief, $\varphi$ holds on all traces of all initial paths.
Note, $\K\varphi$ may hold, although $\varphi$ may not hold on \universeD. 
$\B\varphi$ is defined as $\neg\K\neg\varphi$ and reads as \emph{\enquote{\ego believes that $\varphi$ is possible}}.
Likewise, we define $\K^c$ or $\B^c$ on traces from current states. 
A finite set of BLTL formulae constitutes a \emph{knowledge base} $\kbase\in\kbases$.
A belief \belief satisfies ${\kbase}$, $\belief\models\kbase$, if $\belief\models\varphi$ holds for all $\varphi\in\kbase$.
We assume that \ego's knowledge base varies over time.
Therefore, we extend the labelling function $\Label$ by $\Label_\kbases:\States\rightarrow \kbases$ specifying the available knowledge base at a state. 
Given a history \history, $\kbase_\history$ denotes the knowledge base of $\last(\history)$. 

\begin{example}{A Knowledge Base}\label{ex:kb}
	Let \ego have the following knowledge base $\kbase$ at all states.\\[-5mm]
	{\small
\begin{enumerate}\itemindent-5mm 
		\item { {$\varphi_z=\K \Box ((\neg \xother=5 \land \neg \xother=6) \lor \textsf{undef})$\normalsize}} \hfill\hspace*{2mm}\textit{(\other is at most at $x=4$)}, \\[-6mm]
		\item {{ $\varphi_t=\K \Box (\neg\xego=2 \Rightarrow (\yego=1 \lor \textsf{undef}))$\normalsize}} \hfill \hspace*{2mm}\textit{(a turn may only be possible at $x=2$)},\\[-6mm]
		\item {{ $\varphi_i=\K\, \xego=1$ \hfill \hspace*{2mm}\textit{(\ego starts at $x=1$)}\normalsize}} and \\[-6mm]
		\item {{ $\varphi_{ct}=\K \Box \bigwedge_{t\in\{\textsf{slow, fast}\}} (t \Rightarrow \Next (t\lor undef)) \land (\neg t\Rightarrow \Next ( \neg t\lor \textsf{undef}))$\normalsize}}  \hspace*{-2mm} \textit{(the initial car type does not change)}.\\[-10.5mm]
\end{enumerate}

\normalsize}
\end{example}			
\myparagraph{Belief Formation} %
\Ego updates its beliefs e.g. when it gets new information from its sensors, a clock tick or a message from another agent.
The belief formation function \LabelB captures formally how \ego builds its belief. 

\begin{definition}{belief formation function, knowledge-consistent}\label{def:bform}
	The belief formation function \LabelB, $\LabelB:\histories_{\Obs}\times\kbases\rightarrow \Beliefs$, specifies the belief $\belief$ that \ego derives after perceiving history $\history$ of observations $\Obs\subseteq\Obser$ and while trusting in $\kbase_\history$. 

	We call a belief formation function $\LabelB$ \emph{knowledge-consistent}, if its beliefs satisfy the current knowledge base, i.e., $\LabelB(\history,\kbase_\history)\models\kbase_\history$ for all $\history$.
\end{definition}
Note, that $\Label_\kbases$ is a mean to define a relation between \ego's beliefs and the ground truth, when the belief formation is knowledge-consistent. 
In that case we can for instance specify that the formed beliefs have to be consistent with the last two observations, or with messages of the traffic control system. 

For a history of observations \Small{$h=h_0h_1\ldots h_n$} and a history of knowledge bases \Small{$\kbase=\kbase_0\kbase_1\ldots \kbase_n$}, we denote  by $\LabelBh(h,\kbase)$ the resulting history of beliefs, \Small{$\LabelB(h_0,\kbase_0)\LabelB(h_0h_1,\kbase_1)\ldots\LabelB(h_0h_1\ldots h_n,\kbase_n)$}.
We usually write $\LabelB(\history)$ instead of $\LabelB(\history,\kbase_\history)$. Likewise, we write $\LabelBh(h)$ for $\LabelBh(h,\kbase)$.

\begin{figure}[htbp]
	\centering
	\includegraphics[width=.8\textwidth]{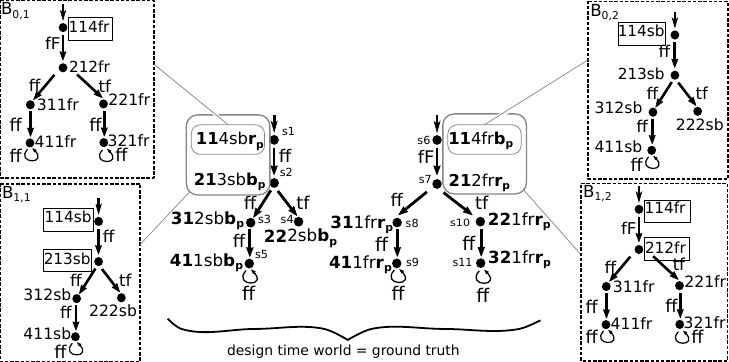}
	\captionof{figure}{Sketch of a belief formation function}\label{fig:bellab}
\end{figure}

\begin{example}{Knowledge-Consistent Belief Formation}\label{ex:knowledgeBelief}
	Let us now see how \ego's belief formation could be like. Let \ego have the knowledge base \kbase of \autoref{ex:kb} and 
	let \ego also be convinced that a red car is fast, while a blue car is slow, i.e., $\varphi_b=\Box (\textsf{fast} \Leftrightarrow \rp) \land \Box (\textsf{slow} \Leftrightarrow \bp)$,
	so all states of \universeD are labelled with $\kbase':=\kbase\cup\varphi_b$ by $\Label_\kbases$.
	In order to satisfy $\kbase$, only beliefs sketched in (a) of \autoref{fig:bel_setup} remain possible.
	The initial belief of a knowledge-consistent belief formation has to consist of the alternative realities depicted in \autoref{fig:init_bel}(a)+(c). 
	When \ego in the real world scenario of Figure \ref{fig:init_bel}(b) incorrectly perceives that \other is blue, it will form $\belief_b$, the belief of Figure~\ref{fig:init_bel}(a).
	$\belief_b$ expresses that \ego thinks it knows that \other is blue and hence a slow car, although it is red in the real world. 
	It moreover captures \ego's expectation of how the future will develop. 
	Similarly, \autoref{fig:init_bel}(c) shows the belief $\belief_r$, which \ego forms, when it is in the real world scenario of \autoref{fig:init_bel}(d) and incorrectly perceives that \other is red.     
	\begin{figure}[htbp]
		\begin{minipage}{0.25\textwidth}
			\centering
			\includegraphics[width=.6\textwidth]{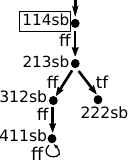}\\
			{(a) \Small{ Alternative reality: Other is slow}}
		\end{minipage}
		\hfill
		\begin{minipage}{0.21\textwidth}
			\centering
			\includegraphics[width=.9\textwidth]{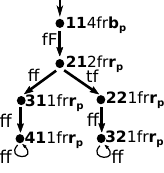}\\[2mm]
			{(b) \Small{Real world: Other is fast}}
		\end{minipage}
		\hfill
		\begin{minipage}{0.25\textwidth}
			\centering
			\includegraphics[width=.6\textwidth]{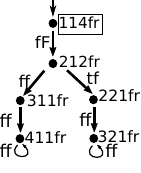}\\
			{(c) \Small{ Alternative reality: Other is fast}}
		\end{minipage}
		\hfill
		\begin{minipage}{0.21\textwidth}
			\centering
			\includegraphics[width=.9\textwidth]{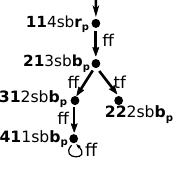}\\[-1mm]
			{(d) \Small{Real world:\\ Other is slow}}
		\end{minipage}
		\caption{The two alternative realities of \ego at its initial state (a)+(c); the real world scenarios (c)+(d) with observable values in bold face type.}\label{fig:init_bel}
	\end{figure}	

	So, since \ego's perception is mistaken, \ego is initially convinced that there is a fast car, when there is a slow car and vice versa.  
	\Ego hence thinks that it should do the turn, when it should not. 
	But \ego does not have to take the decision whether to turn at that time. 
	It first moves one tile forward, while \other simultaneously moves either one or two tiles forward. 
	In both cases, \ego then perceives \other's colour correctly. 

	Let us say, in retrospective \ego is aware of the faulty perception and corrects its belief on \other's car type and colour. 
	It updates the belief to \autoref{fig:second_bel}(a) when it is in the scenario 
	\autoref{fig:init_bel}(b), and to \autoref{fig:second_bel}(b), when in 
	\autoref{fig:init_bel}(d).	
	Figure \ref{fig:bellab} sketches the belief formation so far.
	The observed history \Small{11r$_p$} is mapped to belief \Small{B$_{0,1}$} and \Small{11b$_{p} \mapsto $B$_{0,2}$}, \Small{11r$_p$,21b$_p \mapsto $B$_{1,1}$} and \Small{11b$_{p}$,21r$_p \mapsto $B$_{1,2}$}. 
	The sketched belief formation is knowledge-consistent. 
	Although \ego changed its belief about \other's type, $\varphi_{ct}$ holds for each formed belief. 

	\begin{figure}[htbp]
		\centering
		\begin{minipage}{0.6\textwidth}
			\begin{minipage}{0.32\textwidth}
				\centering
				\includegraphics[width=\textwidth]{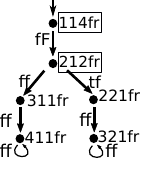}\\
				(a)
			\end{minipage}
			\hfill
			\begin{minipage}{0.29\textwidth}
				\centering
				\includegraphics[width=\textwidth]{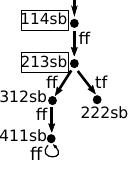}\\
				(b)
			\end{minipage}
			\captionof{figure}{Possible worlds}\label{fig:second_bel}
		\end{minipage}
	\end{figure}	

	Since \ego's beliefs match now the reality, \ego is able to assess the best strategy.
	\Ego can develop its strategy based on its sequence of beliefs, arguing along the lines \enquote{Initially I thought the car is red and fast and that it is a good idea to do the turn. Now I think the car is blue and slow and then the turn is not good idea, since I would collide with other. Since I believe, that my current belief matches the reality, I choose to drive straight on.}  
\end{example}

In the above example, the belief is made up of singletons of alternative realities  only. The next example illustrates the use of several alternative realities.
\begin{example}{Alternative Realities}\label{ex:realities}
	Let us assume that \ego is unsure of its own initial position, thinking that it may initially be at \Small{$x=1$} or \Small{$x=2$}, as sketched in \autoref{fig:bel_setup}(a)+(c). 
	So \ego deems two realities possible, when at state $s1$. 
	In both realities \other is red, but in one reality, $r_1$, \ego is at \Small{$x=1$}, in the other, $r_2$, at \Small{$x=2$}. Similarly, \ego deems two realities possible when at $s6$.
	\begin{figure}[h]
		\centering
		\includegraphics[width=.8\textwidth]{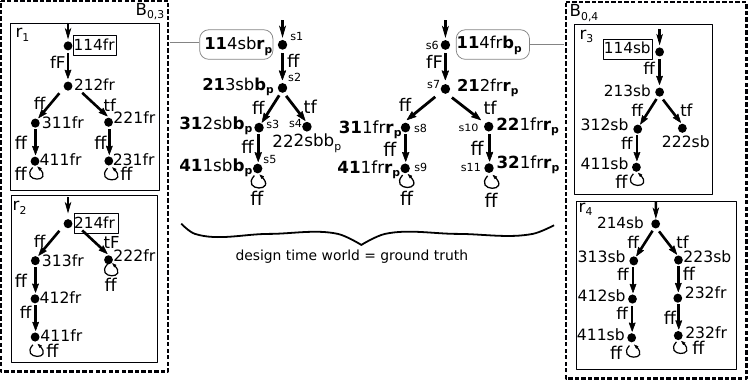}
		\caption{Sketch of a belief formation function when \ego is unsure of its initial position}\label{fig:bellab3}
	\end{figure}
\end{example}

\myparagraph{Doxastic Model} 
To summarize,  a doxastic model \epi is given  by a tuple (\universeD,\goalList,\kbases,\Obs,\Beliefs,\LabelB) of the design-time world \universeD, the prioritized list of goals \goalList, the knowledge bases \kbases, a set of observations $\Obs$, the set of possible beliefs \Beliefs of a \HAS and a belief formation \LabelB. 
The belief formation describes how a system links observations made within the world \universeD to the \HASes inner representation of the world, i.e. the beliefs \Beliefs that a \HAS can possibly build.
The world \universeD is considered as ground truth during the design. 
Later design steps have to take care of the gap between \universeD and the real world.

\section{Autonomous Decisions}\label{sec:autonom}
%
\normalsize
In this section, we formalize what it means that \ego decides autonomously. 
\Ego decides autonomously, if its decision is the result of \ego's simulative assessment of its actions on content of its beliefs and the decision is sensible wrt \ego's observations. 

Our formalization hence distinguishes between autonomous systems --systems that take decisions based on the belief content-- and automatic systems --systems that play out rule-based decisions that have been taken at design time.
In order to formalize \emph{autonomous-decisive} we contrast (i) strategies that have access to ground-truth with (ii) strategies that can only observe the formed beliefs, and (iii) strategies that run as simulation within the beliefs. 
The different strategy notions are introduced step by step in the sequel and an overview of the notions is given in \autoref{tab:nutshell}.
\myparagraph{Truth-Observing and Doxastic Strategies} 
In our framework, we use truth-observing strategies as reference of what would be achievable, if \ego could directly access the ground truth \universeD via a set of propositions \props. 
To this end, we say \ego implements a \emph{$\props$-truth-observing strategy} $\strats:(2^{\props})^+\rightarrow\Act_{\ego}$, if \ego chooses its actions based on the history of values of $\props\subseteq\Props_{d}$ as observed in \universeD. 
Implementing \strats means that \ego decides what to do at a state $\Path(i)$, reached via the path \Path in \universeD, based on $\Label(\Path_{\leq i})|_{\props}$, i.e. based on the true past values of $\props$ observed thus far.
A truth-observing strategy \strats together with a sequence of environment actions $\e\in\Act_{\env}^{\omega}$ determines a set of traces, $\Cmp(\e,\strats)$.
Formally, $\Cmp(\e,\strats) = \{\cmp_0 \cmp_1\ldots \in (2^{\Props_D})^\omega |
\exists \textit{ path } \Path \textit{ from } \Init_D, \forall i\geq 0:
\cmp_i = \Label_D(\Path(i)) \land 
\act_i:=\strats(\Label_D(\Path_{\leq i})|_{\props}) \land
(\act_i,\e(i))\in\Label_D(\Path_i,\Path_{i+1})\}$.

Since \ego has no direct access to the ground truth,  we also formalize that \ego decides based on the history of beliefs.
At a state $\Path(i)$ in \universeD \ego takes a decision based on the history of its beliefs $\bhist_0\ldots\bhist_i$ that \ego has built along $\Path_{\leq i}$.
\Ego implements a doxastic strategy $\stratb:\Beliefs^{+}\rightarrow\Act_{\ego}$ on $\universeD$, if \ego chooses its actions based on the history of its beliefs.
A strategy \stratb together with a  sequence of environment actions $\e\in\Act_{\env}^{\omega}$ determines a set of traces in \universeD, just like for truth-observing strategies. 
The set of traces is $\Cmp(\e,\stratb) = \{\cmp_0 \cmp_1\ldots \in (2^{\Props_D})^\omega |
\exists \textit{ path } \Path \textit{ from } \Init_\epi, \forall i\geq 0:
\cmp_i = \Label_\epi(\Path(i)) \land 
\act_i:=\stratb(\LabelB(\Path_{\leq i})) \land
(\act_i,\e(i))\in\Label_\epi(\Path_i,\Path_{i+1})\}$.
Note that a doxastic strategy also depends on observations of the doxastic model \epi due to the belief formation \LabelB (cf. \autoref{def:bform}).

Since truth-observing and doxastic strategies both determine traces for a given sequence of environment actions,  we can compare them straight forwardly:  \Ego achieves \goalList up to $n$ on \universeD by strategy \strat, if for all $\e\in\Act_{\env}^{\omega}$ all $\cmp\in\Cmp(\e,\strat)$ satisfy $\goalList$ up to $n$. 
A strategy \emph{\strat  \goalList-dominates $\strat'$ on $\universe$}, $\strat'\leq_{\universe}\strat$, iff $\strat'$ achieves \goalList up to $m$ and \strat up to $k\geq m$. 

\begin{example}{Truth-Observing and Doxastic Strategies}
	As an example of a dominant \props-truth-observing strategy, let us consider the goal list of \autoref{ex:goals}, the world model in \autoref{fig:setup}(b), $\props:=\{\xego,\textsf{slow},\textsf{fast}\}$ and the strategy \strats that chooses to drive straight on, if \other is fast, and it chooses to turn, if \other is slow
	(it maps \Small{$1s\mapsto f$, $1s,2s\mapsto f$, $1s,2s,3s \mapsto f$, $1s,2s,3s,4s\mapsto f$, and  $1f\mapsto f$, $1f,2f\mapsto t$, \ldots}). 
	Strategy \strats achieves \goalList up to $2$ and is a dominant ($\props$-truth-observing) strategy, since in all cases collision-freedom is guaranteed and
	in case the car is slow, no other strategy is able to realize both, the turn and collision freedom.

	Let us now define a doxastic strategy \stratb for a belief formation, as sketched in \autoref{fig:bellab}.
	Let \stratb be a doxastic strategy with \Small{$B_{0,1}\mapsto f$, $B_{0,1}\,B_{1,1}\mapsto f$, $\ldots$} and \Small{$B_{0,2}\mapsto f$, $B_{0,2}\,B_{1,2}\mapsto t$, $\ldots$}. 
	Just like \strats,  \stratb chooses to turn when \other is fast (\Small{$B_{0,2}\,B_{1,2}\mapsto t$}), and it chooses to drive straight on, if \other is slow (\Small{$B_{0,1}\,B_{1,1}\mapsto f$}). 
	As there is \enquote{no better} strategy, \stratb is dominant. 
\end{example}

Next, we want to capture that \ego chooses its actions based on the \emph{content} of its beliefs. In order to motivate our formalization, let us consider the following example.
\begin{example}{Decisions Not Based on the Belief Content}
	We modify our running example slightly: Let us assume the colour perception is severely broken and permanently switches red to blue and vice versa. 
	In \autoref{fig:bellab2} the changed world model is given, and a belief formation is sketched, that derives a belief that trusts in the colour perception, i.e., if the sensors say the other car is red, then \ego believes the other car is red.
	\begin{figure}[h]
		\centering
		\includegraphics[width=.8\textwidth]{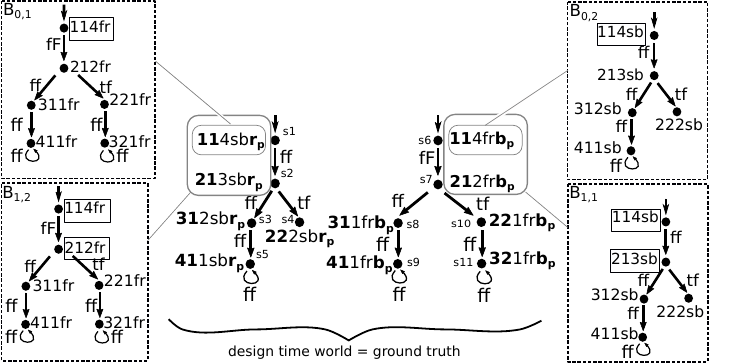}
		\caption{Sketch of a belief formation function for severely broken colour perception}\label{fig:bellab2}
	\end{figure}

	Let $\stratb'$ be a doxastic strategy with \Small{$B_{0,1}\mapsto f$, $B_{0,1},B_{1,2}\mapsto f$, $\ldots$} and \Small{$B_{0,2}\mapsto f$, $B_{0,2},B_{1,1}\mapsto t$, $\ldots$}. 
	Just as \strats and \stratb,  the strategy $\stratb'$ realizes a turn on \universeD, if \other is fast (\Small{$B_{0,2},B_{1,1}\mapsto t$}), and \ego drives straight on, if \other is slow (\Small{$B_{0,1},B_{1,2}\mapsto f$}). 
	So $\stratb'$ is also dominant, but $\stratb'$ makes no sense from \ego's perspective. 
	In case \other is fast, \ego believes that \other is slow, since it trusts its sensors.
	In this case, \ego extrapolates that doing the turn  would cause a collision, 
	but according to $\stratb'$, \ego chooses to take the turn.
	Vice versa,  $\stratb'$ chooses to drive straight on, when \ego believes \other is fast and although \ego extrapolates that taking the turn is alright. 
\end{example}
%
\myparagraph{Autonomous-decisive System}
We now formalize what it means that \ego decides based on the content of its belief. 
\Ego always chooses an action that belongs to a strategy that is dominant in \ego's current belief \belief.
To capture this formally, we introduce the notion of \emph{possible-worlds strategy}. 
A \emph{possible-worlds strategy}  is a function $\stratw:{(2^{\Props_\belief})^{+}}\rightarrow\Act_\ego$ and it is simulated on the alternative realities of \ego's current belief \belief. 
This results in \enquote{believed traces}.
We define this set of traces in an alternative reality $\reality=(\world,\cStates)\in\belief$ for a (believed) sequence of environment actions $\e\in\Act_{\env}^{\omega}(\world)$ as 
$\Cmp(\e,\stratw,\reality) = \{\cmp_0 \cmp_1\ldots \in (2^{\Props})^\omega | \exists \textit{ path } \Path\textit{ in }\world\textit{ from }\Init:  \forall i\geq 0:
\cmp_i = \Labelw(\Path(i)) \land 
\act_i:=\stratw(\Labelw(\Path_{\leq i})) \land
(\act_i,e(i))\in\Labelw({\Path}_{i},\Path_{i+1}) \}$.
We generalize the notion of \goalList-dominance to possible-worlds strategies. 
A possible-worlds strategy \stratw \goalList-dominates a possible-worlds strategy $\stratw'$ in \belief, if \stratw \goalList-dominates $\stratw'$ in all realities $\reality\in\belief$.
\begin{example}{Possible-Worlds Strategy}\label{ex:pws}
	Consider the possible-worlds strategy that chooses to turn, if \other is fast, and to drive straight on, if \other is slow, i.e., consider \stratw with  \Small{$114fr\mapsto f$, $114fr, 212fr\mapsto t$,  $114fr, 212fr,221fr\mapsto f$, \ldots, $114sb\mapsto f$, $114sb,213sbb_p\mapsto f$, \ldots}  
	In the belief $B_{0,1}$ of \autoref{fig:bellab2} \ego thinks that it is at the initial state \Small{$114fr$}. \Ego follows \stratw by choosing \Small{$\stratw(114fr)=f$}. 
	In contrast, the belief $B_{1,2}$ captures that \ego thinks to have already made one move and is now at state \Small{$212fr$}.  \Ego follows \stratw by choosing  \Small{$\stratw(114fr,212fr)=t$}.
\end{example}

\begin{table}
	\FramedBox{4.8cm}{0.97\textwidth}{
		\small
		Strategy tyes:\\[-6mm]
		\begin{itemize}
			\item\small \emph{truth-observing strategy} $\strats:(2^{\props})^+\rightarrow\Act_{\ego}$\\ 
				observes the  ground truth world $\universeD$ via $\props\subseteq\Props_{\universeD}$ and takes decisions based on their history; serves as comparative reference of what is achievable given $\props$ could be observed directly\\[-6mm]
			\item\small \emph{doxastic strategy} $\stratb:\Beliefs^{+}\rightarrow\Act_{\ego}$\\
				observes the beliefs to take decisions and takes decisions based on their history; represents the decision making of autonomous and automatic systems\\[-6mm]
			\item\small \emph{possible-worlds strategy} $\stratw:{(2^{\Props_\belief})^{+}}\rightarrow\Act_\ego$\\
				captures how a \HAS \enquote{simulates} its strategies within the alternative realities; decisions are taken based on the believed history within the respective alternative reality\\[-6mm]
		\end{itemize}
		A strategy \strat  \emph{\goalList-dominates} $\strat'$, $\strat'\leq\strat$, iff $\strat'$ achieves \goalList up to $m'$ but \strat achieves \goalList up to $m'$ with  $m'\leq m$.\\
		\normalsize
		}
		\caption{Strategy types \& dominance in a nutshell}\label{tab:nutshell}
\end{table}

A peculiarity of possible-worlds strategies is, that they can be indecisive for a belief:
A belief may contain alternative realities encoding even contradictory information. E.g. let us assume that \ego is currently not sure where its goal position is. In the alternative reality $\reality_1$ its goal position is to its left while in $\reality_2$ the goal position is right of \ego. \Ego's strategy is to move to the left, when in $\reality_1$ and to move to the right when in $\reality_2$. Since \ego deems both realities possible, it cannot decide whether to turn right or left. 

\begin{definition}{current-state decisive possible-worlds strategy, current-state choice}\label{def:cdec}
	A possible-worlds strategy \stratw is \emph{current-state decisive in a belief \belief}, if \stratw specifies one and the same action for all the believed current states of all alternative realities of this belief, i.e.\\
	{$\exists\, \act_{\belief}\in\Act_\ego$ such that $\forall \reality=(\world,\cStates)\in\belief: \forall \textit{ paths }\pi\textit{ in }\world: \pi(0)\in\Init\land\textit{last}(\pi)\in\cStates\Rightarrow\stratw(\pi)=\act_\belief$}.
	Let \Small{$\Act(\stratw,\belief)$} denote the action \Small{$\act_\belief$} that $\stratw$ chooses at the current states {$\bigcup_{(.,\cStates)\in\belief}\cStates$} of \belief. We call \Small{$\Act(\stratw,\belief)$} be the current-state choice of \stratw.

\end{definition}

\begin{sloppypar}
We can now define our notion of autonomous-decisive system. %
We consider a  system to be autonomous-decisive, if \Small{(i)} it takes its decisions based on the content of its beliefs and \Small{(ii)} the resulting behaviour is sensible. 
We capture \Small{(i)} by requiring that  there is a doxastic strategy $\stratb$, i.e., a strategy that can only observe the beliefs.
Moreover, $\stratb$ takes only  actions that seem the best choice according to the simulation within its beliefs.
With other words, $\stratb$ only takes actions that are the current-state choice of a dominant, current-state decisive possible-worlds strategy.
We capture \Small{(ii)} by requiring that $\stratb$ does not perform worse than a strategy that decides based on the made observations, i.e. $\stratb$ is not $\goalList$-dominated by a \Obs-truth-observing strategy. We hence require that the belief formation captures the gist of observations.
\end{sloppypar}
\begin{definition}{doxastic system, autonomous-decisive system}\label{def:autonom}
	Let a doxastic model \\$\epi=(\universeD,\goalList, \kbases,$ $\Obs, \Beliefs, \LabelB)$ be given. 
	
	A doxastic strategy $\stratb$ is \emph{autonomous-decisively} achieving $\psi$ up to $n$, 
	if $\stratb$ achieves $\psi$ up to $n$ and it always follows dominant, current-state decisive possible-worlds strategies, i.e.,\\ 
	for all observed histories $h$, it holds that there is a dominant possible-worlds strategy $\stratw$ with {$\stratb(\LabelBh(\history))=\Act(\stratw,\LabelB(\history))$} that is current-state decisive in {$\LabelB(\history)$}.

	A \emph{(doxastic) system} is a pair $\sys=(\epi,\stratb)$ of a doxastic model $\epi$ and a doxastic strategy $\stratb$.  
	\sys is an \emph{autonomous-decisive system}, 
	if there is an $n$ such that $\stratb$ autonomous-decisively achieves $\psi$ up to $n$ and 
	$\stratb$ is not $\psi$-dominated by a \Obs-truth-observing strategy.
\end{definition}

\begin{example}{Autonomous, Non-Autonomous,  Automatic}
\begin{sloppypar}
	Let us consider an example of an autonomous \ego in the  setting of \autoref{fig:bellab}, where the sensor only initially switches colours, and consider the possible-worlds strategy \stratw of \autoref{ex:pws} (turn, if \other is fast, and drive on, if \other is slow). 
	When \ego initially evaluates its situation in $s_1$ of \universeD, it believes that the situation is as described by {$B_{0,1}$}, i.e. \other is a fast, red car.
	\Ego can decide to follow \stratw in {$B_{0,1}$}, as it seems a good choice -- \stratw is dominant and current-state decisive in {$B_{0,1}$}. 
	According to its extrapolation, it would move one step forward, and then it would successfully take the turn.
	After actually moving forward, \ego evaluates the situation in $s_2$.
	In $s_2$ \ego believes in {$B_{1,1}$} reflecting that \ego now truthfully perceives \other's colour as blue (cf. \autoref{ex:knowledgeBelief}).  
	Again, \stratw is a dominant current-state decisive possible-worlds strategy and determines \textsf{f} as the next move.
	Along this line, it is easy to see that \ego can implement a doxastic strategy \stratb that chooses the action that \stratw determines for the respective $\LabelB(h)$.
\end{sloppypar}

	For an example of a non-autonomous \ego, we modify our running example slightly.
	Let us assume that \ego is unsure of its own initial position, thinking that it may initially be at \Small{$x=1$} or \Small{$x=2$}, as sketched in \autoref{fig:bellab3}. 
	Let a belief formation be given that evolves the initial beliefs {$B_{0,3}$} and {$B_{0,4}$} analogously to \autoref{ex:knowledgeBelief}, that is, \ego perceives the correct colour after moving one step forward.
	The possible-worlds strategy \stratw is still a dominant strategy but not current-state decisive, since for example in {$B_{0,3}$} \ego would do the turn at $s1$ due to reality $r_2$, and it would also drive straight on due to $r_1$. Hence, there is no dominant possible-worlds strategy in {$B_{0,3}$} that is able to determine one action. \Ego cannot decide autonomously.

	Note that we nevertheless can specify a dominant doxastic strategy \stratb for this case, but its actions are not chosen based on the belief content: \Ego chooses to turn after one step when its initial belief was {$B_{0,4}$} (\Small{$B_{0,4},B_{1,4}\mapsto t$}), otherwise it drives straight on.
	This strategy is dominant and could be used to build an \emph{automatic} system, where \ego just plays out \stratb. Such a strategy might be useful when an engineer knows that \ego will start from \Small{$x=1$} but did not equip \ego with this information.
\end{example}

A system that is not autonomous-decisive cannot determine by itself which action is currently appropriate. A goal for the  design of a \HAS is to ensure that a system is autonomous-decisive.
\begin{theorem}[Autonomous Decisiveness]\label{th:auto}
	Let  $\epi=(\universeD,\goalList,\priorK,\Obs,\Beliefs)$, a doxastic model without belief formation function, be given.

	We can decide whether there is a knowledge-consistent belief formation \LabelB and a  doxastic strategy \stratb such that $\sys=(\epi,\LabelB,\stratb)$ is an autonomous-decisive system that achieves $\goalList$ up to $n$. 
	If such \LabelB and \stratb exist, we can synthesize them. 
\end{theorem}

\begin{proofsketch}\label{proof}
	We build a Kripke structure $\universeD'$ such any truth-observing strategy \strats in $\universeD'$ encodes a belief formation $\LabelB$ and a doxastic strategy \stratb, such that \LabelB is knowledge-consistent and \stratb always follows dominant, current-state decisive possible-worlds strategies, and if \strats achieves \goalList up to $n$, also $\stratb$ does.

	We first determine the set of actions $\Act(\belief)$ that an autonomous-decisive system can choose when currently having the belief \Small{$\belief=\{\reality_1,\ldots,\reality_n\}\in\Beliefs$}. 
	An action \act can only be chosen by a doxastic strategy of an autonomous-decisive system, if it is chosen by a dominant, current-state decisive possible-worlds strategy \stratw for the current states of \belief (cf.~\autoref{def:autonom}). 
	Note that there may be several dominant possible-worlds strategies for \belief. 
	We build a single reality $\reality_{\belief}$ by the disjoint union of all alternative realities \Small{$\reality_{\belief}:=((
	\dbigcup_{r_i}\{\States_i\},
	\dbigcup_{r_i}\Edges_i,
	\dbigcup_{r_i}\Label_i,
	\dbigcup_{r_i}\Init_i),
	\dbigcup_{r_i}\{{\cStates}_i\})$}.
	If necessary, we can make the realities disjoint by renaming their states but keeping their structure.
	In order to judge how well a current-state decisive, possible-worlds strategy has to perform in order to be dominant, we determine the maximal $\mn$ up to which $\psi$ can be achieved by any possible-worlds strategy. 
	Therefore, we check whether we can synthesize a strategy in $\reality_{\belief}$ that achieves $\psi$ up to $\mn=n$ down to $\mn=0$, if necessary \cite{LTLSynth}.  
	Note that in the worst case only the trivial goal \true, i.e. $\mn=0$, can be achieved.
	In order to check whether \act is chosen by a dominant, current-state decisive strategy, 
	we first check whether \act is enabled at all current states of {$\reality_{\belief}$}.
	If \act is enabled, we make $\act$ the only choice at the current states. 
	Therefore, we build the modified Kripke structure {$\reality_{\belief,\act}$}.
	We replace all arcs from the current states of $\reality_{\belief}$  that are not labelled with \act by arcs leading to \sundef. 
	If we can synthesize a strategy {$\reality_{\belief,\act}$} that achieves $\goalList$ up to $\mn$, \act is a current-state choice of that current-state decisive, possible-worlds strategy.
	By iterating over all actions, we collect the set of the actions {{\Act}(\belief)}. 

	In order to judge how well the doxastic strategy \stratb has to perform, we determine the maximal $\mn$ up to which $\psi$ can be achieved by any \Obs-truth-observing strategy in \universeD. 
	If $\mn<n$, then there is no (\LabelB,\stratb) such that \sys achieves \goalList up to $n$. Hence let $n\leq \mn$.
	We build the modified Kripke structure $\universeD'$ as follows. 
	If $\universeD$ has an $\act$-labelled arc from a state $\state$ labelled, we keep it only if there is a belief $\belief\in\Beliefs$ with $\act\in{\Act(\belief)}$ and $\belief\models\Label_\kbases(\state)$~\cite{MC}. 
	Otherwise, we redirect the arc to \sundef.
	We synthesize  a dominant strategy $\strats$ on {$\universeD'$}~\cite{LTLSynth}.
	We derive the belief formation {$\LabelB$} from $\strats$ by $\LabelB(\history):=\belief$ for a belief $\belief$ with $\strats(\history)\in\Act(\belief)$. 
	We define \stratb to be the doxastic strategy defined by $\stratb(\LabelBh(\history)):=\strats(\history)$.
	In case $\strats$ achieves $n$, \sys is an autonomous-decisive system. 
	Otherwise the truth-observing strategies perform better.  
\end{proofsketch}

\section{Related and Future Work}\label{sec:relwork}
%
Epistemology is the theory of knowledge and concerned with information-processing and cognitive success \cite{CollinsEpi, Sheffield}.
Doxastic means \enquote{relating to belief}~\cite{Collins}. 
By using the term \enquote{doxastic}, we want to stress that our formalism focuses on beliefs.
Note in particular that an alternative reality can vary arbitrarily from the ground truth \universeD, and a \HAS does not necessarily believe that the ground truth is even possible. 
In the epistemic logic literature, the semantics of doxastic languages are often given via \emph{doxastic models}, that are special Kripke structures \cite{EpistemicsLogics}.
A doxastic model $(S,v,\rightarrow_i)$ consists of a set of nodes $S$ representing possible worlds $w$, a valuation function $v: S\rightarrow 2^{\Props}$ for the set of atomic facts \Props and a belief relation $\rightarrow_i$  for each player $i$, that specifies \enquote{$i$ deems $w'$ possible in $w$} if \enquote{$w$ $\rightarrow_i$ $w'$}. With other words, the belief of $i$ at $w$ is defined as the worlds accessible via the agent $i$'s belief relation, $\rightarrow_i$ \cite{EpistemicsLogics,DoxasticMeyer2003}. 
In this paper, we use complex possible worlds instead of the plain nodes of a Kripke structure. 
Our alternative realities encode the believed histories, the current states and possible futures.
A \HAS uses alternative realities to simulate its strategy in order to decide on its current action. 
In our framework, a reality constitutes an extensive form two-player zero-sum game, where the winning condition is defined by the list of LTL goals.
While the possible-world strategies have perfect information on the beliefs, the belief formation, which we also regard as a strategy, is based on partial observations and the currently available knowledge/hard beliefs.

A couple of epistemic temporal logics have been suggested for specifying aspects of knowledge throughout time for multi-agent systems. These logics combine temporal logics with knowledge operators, like KCTL \cite{CTLKMC}, KCTL$^*$ or HyperCTL$^{*}_{lp}$~\cite{HyperCTL}. They are interpreted over Kripke structures. 
But since an agent $i$ has its local view, only certain propositions are assumed to be observable, so that an observational equivalence relation $\sim_i$ on the traces arises. 
\enquote{Agent $i$ knows $\varphi$} then means that $\varphi$ holds on all $i$-equivalent initial traces. 
The alternating time temporal logics (ATL)~\cite{ATL} has been developed for reasoning about what agents can achieve by themselves or in groups throughout time. 
In ATL, the path quantifiers of CTL are replaced by modalities that allow to quantify paths in the control of groups of agents.
ATL is interpreted over concurrent game structures~(CGS), which are labelled state transition systems. 
By adding a knowledge operator, ATL has been extended to an epistemic variant, ATEL~\cite{ATEL}. 
To this end the concurrent game is extended by an observational equivalence relation per agent modelling the agent's limited view.
In contrast to be above logics, a BLTL formula is interpreted on a belief \belief, i.e. a set of alternative realities. 
Since the set of possible beliefs \Beliefs is finite, a formula $\K\varphi$ means the finite conjunction $\bigwedge_{\reality\in\belief}\reality\models\varphi$.  
But just like the logics above, we assume that a \HAS can only partially observe the ground truth -- our doxastic strategies cannot distinguish between histories that are observational equivalent. 
Our beliefs, however, cannot straightforwardly be expressed in terms of an equivalence on the ground truth, since an alternative reality may be a distinct Kripke structure and a belief does not have to include the ground truth.

The field of epistemic planning is concerned with computing plans (\enquote{a finite succession of events} \cite{DELStrat}) that achieve the desirable state of knowledge from a given current state of knowledge~\cite{GIntroDEL}.  
DEL, dynamic epistemic logic, is a formalism to describe planning tasks succinctly by a semantic and action model based approach. 
Epistemic models capture the knowledge state of the agents, and epistemic action models describe how these are transformed. 
An evolution results from  a stepwise application of the available actions. 
In \cite{DELStrat} distributed synthesis of observational-strategies for multiplayer games are considered.  
While ATEL and DEL allow for reasoning about a combination of knowledge and strategies, we are interested in the belief \emph{formation}. 
We ask whether there exists a belief formation that justifies a strategy that successfully achieves temporal goals within a given ground truth world. 

Properties of belief formation are studied in the field of belief revision and update. 
Belief revision is done when a new piece of information contradicts the current information, and it aims to determine a consistent belief set. 
Belief updates may be necessary when the world is dynamic~\cite{BelRevIntro}. 
The works in this field are concerned with rational belief formation, following e.g. some guiding principle like making minimal changes~\cite{BelRevIntro}. 
In our work, we consider very general belief formation functions, since we focus on \HASes.
The belief of a \HAS may be determined by a composition of different components, and there may not necessarily be an entity that ensures that the resulting belief is rational. 
We imagine that during the design, requirements regarding the rationality might be specified. 
So, approaches examining whether an appropriate belief formation exists might be valuable tools for the development of \HASes. 

BDI agents are rational agents with the mental attitudes of belief~(B), desire~(D) and intention~(I)~\cite{BDI}.
Beliefs describe what information the agent has, 
desires represent the agent's motivational state and specify what the agent would like to achieve, 
while intentions represent the currently chosen course of action. 
These attitudes allow an agent balancing between deliberation about its course of action and its commitment to the chosen course of action.
In our framework, an agent deliberates about its course of action at each state. 
We do not enforce commitment to a certain course of action, as we are interested in whether some belief formation exists.
Nevertheless, the framework conceptually allows capturing notions of commitment, and we plan to examine these in future work. 
Basically, a chosen action represents a set of believed best possible world strategies. These can be considered as the current intent. 
So, a notion of commitment could require that (some) strategies of the previous belief are still best strategies in the current belief.
An engineer may then specify when a \HAS should be committed.

In this paper, we call a system autonomous-decisive, if it can choose its actions based on its belief content, i.e.,  when it always chooses actions that it believes can be extended to a best strategy.
The concept that an agent adjusts its actions following the believed most promising strategy is certainly not new. 
For instance in \cite{motion}, Wang et al. present an approach to motion planning for autonomous cars in competitive scenarios.
\Ego solves iteratively several rounds of trajectory optimization, whose fixed point represents a Nash equilibrium in the trajectory space of all cars.  
In \cite{ComGame}, a two player game with communication is considered, and a generic framework is defined that uses belief revision techniques to take the communication into account. Player B makes announcements and then player A revises its preferences by analysing what its best response would be.

Although we motivated our term of \emph{autonomous-decisive} by referring mostly to \HASes,  we think the term is not limited to systems of a high level of autonomy, like level {3} to {5} of the norm {SAE J3016}.
Our notion characterizes how a system takes decisions, while the SAE levels classify the interaction between driver and system.
A lot of works characterize notions of autonomous systems.
For instance, \cite{Autonomics} focuses on the challenges of a foundation for autonomous systems. 
According to the authors, autonomous behaviour embodies five autonomous behavioural functions: perception, model update, goal management and planning and self-adaptation. 
They further state  that decision-making is a
central aspect and at the very heart of autonomous system
engineering. 
In this paper, we clearly focus on decision-making.
The paper  \cite{Autonomous} explores the intelligence and system foundations of autonomous systems. 
According to the authors, autonomy aggregates hierarchically from reflexive, imperative and adaptive intelligence. 
They consider autonomy as a property that enables a system to change their behaviour in response to unanticipated events during operation (thereby following \cite{AutoWat}) and \enquote{without human intervention} (following \cite{IntelAdSys}).
Our notion of autonomous-decisiveness ensures that no human intervention is necessary for the decision-making. 
As our \ego can only partially observe the world, one might argue that behaviour is unexpected, when observations of the ground truth do not match with the extrapolation of prior beliefs. 
When \ego adjusts its beliefs and adopts its behaviour, it hence autonomously adjusts to unexpected events.
We rather envision a different research emphasis to approach his topic.
In order to characterize \ego's ability to adjust to unexpected events, we envision to develop notions of strategic/belief formation robustness that specify when an application domain differs too much from the envisioned design time domain to allow \ego taking autonomous decisions.

Moreover, we plan to use the framework to tackle the question \enquote{What is relevant for the \HAS?}, capturing the trade-off between observations and knowledge when the belief formation is given.

\section{Conclusion}\label{sec:concl}
In this paper, we are concerned with safety critical systems that autonomously  assess their situation and decide what to do based on this assessment. 
Such \HASes operate in overwhelmingly complex environments and have to deal with partial observability and distorted perceptions. 
They construct an inner representation of the world that is based on insights of the world and develops over time, when new observations are made. 
We develop a doxastic framework in order to formalize, what it means that such a system autonomously takes decisions.
In contrast to automatic systems, that decide based on their beliefs how to play out pre-determined strategies, a system decides autonomously, if it chooses an action that is justified by extrapolation of the system's beliefs. 
Given a prioritized list of LTL goals, a given design context modelled as Kripke structure and given it has been decided on the build-in knowledge and the system's possible beliefs, it is decidable whether a belief can be formed that enables a system to autonomously achieve its goals and we can synthesize a witness.
We consider this work as a start to examining belief formation with respect to \HAS's strategic success. 
\paragraph*{Acknowledgements}
I like to thank the anonymous reviewers for their valuable comments.

\bibliographystyle{eptcs}
\bibliography{refs}

\begin{thebibliography}{10}
\providecommand{\bibitemdeclare}[2]{}
\providecommand{\surnamestart}{}
\providecommand{\surnameend}{}
\providecommand{\urlprefix}{Available at }
\providecommand{\url}[1]{\texttt{#1}}
\providecommand{\href}[2]{\texttt{#2}}
\providecommand{\urlalt}[2]{\href{#1}{#2}}
\providecommand{\doi}[1]{doi:\urlalt{http://dx.doi.org/#1}{#1}}
\providecommand{\bibinfo}[2]{#2}

\bibitemdeclare{article}{ATL}
\bibitem{ATL}
\bibinfo{author}{Rajeev \surnamestart Alur\surnameend},
  \bibinfo{author}{Thomas~A. \surnamestart Henzinger\surnameend} \&
  \bibinfo{author}{Orna \surnamestart Kupferman\surnameend}
  (\bibinfo{year}{2002}): \emph{\bibinfo{title}{Alternating-Time Temporal
  Logic}}.
\newblock {\sl \bibinfo{journal}{J. ACM}}
  \bibinfo{volume}{49}(\bibinfo{number}{5}), p. \bibinfo{pages}{672–713},
  \doi{10.1145/585265.585270}.

\bibitemdeclare{inproceedings}{ComGame}
\bibitem{ComGame}
\bibinfo{author}{Guillaume \surnamestart Aucher\surnameend},
  \bibinfo{author}{Bastien \surnamestart Maubert\surnameend},
  \bibinfo{author}{Sophie \surnamestart Pinchinat\surnameend} \&
  \bibinfo{author}{Fran{\c{c}}ois \surnamestart Schwarzentruber\surnameend}
  (\bibinfo{year}{2015}): \emph{\bibinfo{title}{Games with Communication: From
  Belief to Preference Change}}.
\newblock In: {\sl \bibinfo{booktitle}{PRIMA 2015: Principles and Practice of
  Multi-Agent Systems}}, \bibinfo{publisher}{Springer}, pp.
  \bibinfo{pages}{670--677}, \doi{10.1007/978-3-319-25524-8\_50}.

\bibitemdeclare{inbook}{PrinciplesOfMC}
\bibitem{PrinciplesOfMC}
\bibinfo{author}{Christel \surnamestart Baier\surnameend} \&
  \bibinfo{author}{Joost-Pieter \surnamestart Katoen\surnameend}
  (\bibinfo{year}{2008}): \emph{\bibinfo{title}{Principles of Model Checking}},
  chapter~\bibinfo{chapter}{5}, pp. \bibinfo{pages}{229--239}.
\newblock \bibinfo{publisher}{The MIT Press}.

\bibitemdeclare{article}{CTLKMC}
\bibitem{CTLKMC}
\bibinfo{author}{Mario \surnamestart Benevides\surnameend},
  \bibinfo{author}{Carla \surnamestart Delgado\surnameend},
  \bibinfo{author}{Carlos \surnamestart Pombo\surnameend},
  \bibinfo{author}{Luis \surnamestart Lopes\surnameend} \&
  \bibinfo{author}{Ricardo \surnamestart Ribeiro\surnameend}
  (\bibinfo{year}{2008}): \emph{\bibinfo{title}{A Compositional Automata-based
  Approach for Model Checking Multi-Agent Systems}}.
\newblock {\sl \bibinfo{journal}{Electronic Notes in Theoretical Computer
  Science}} \bibinfo{volume}{195}, pp. \bibinfo{pages}{133--149},
  \doi{10.1016/j.entcs.2007.08.030}.
\newblock \bibinfo{note}{Proc. of the Brazilian Symposium on Formal Methods
  (SBMF 2006)}.

\bibitemdeclare{inproceedings}{GIntroDEL}
\bibitem{GIntroDEL}
\bibinfo{author}{Thomas \surnamestart Bolander\surnameend}
  (\bibinfo{year}{2017}): \emph{\bibinfo{title}{A Gentle Introduction to
  Epistemic Planning: The {DEL} Approach}}.
\newblock In: {\sl \bibinfo{booktitle}{Proc. of the 9th Workshop on Methods for
  Modalities, M4M@ICLA 2017}}, {\sl \bibinfo{series}{{EPTCS}}}
  \bibinfo{volume}{243}, pp. \bibinfo{pages}{1--22}, \doi{10.4204/EPTCS.243.1}.

\bibitemdeclare{inproceedings}{HyperCTL}
\bibitem{HyperCTL}
\bibinfo{author}{Laura \surnamestart Bozzelli\surnameend},
  \bibinfo{author}{Bastien \surnamestart Maubert\surnameend} \&
  \bibinfo{author}{Sophie \surnamestart Pinchinat\surnameend}
  (\bibinfo{year}{2015}): \emph{\bibinfo{title}{Unifying Hyper and Epistemic
  Temporal Logics}}.
\newblock In: {\sl \bibinfo{booktitle}{Foundations of Software Science and
  Computation Structures}}, \bibinfo{publisher}{Springer}, pp.
  \bibinfo{pages}{167--182}, \doi{10.1007/978-3-662-46678-0\_11}.

\bibitemdeclare{book}{MC}
\bibitem{MC}
\bibinfo{author}{Edmund~M \surnamestart Clarke\surnameend},
  \bibinfo{author}{Orna \surnamestart Grumberg\surnameend} \&
  \bibinfo{author}{Doron~A. \surnamestart Peled\surnameend}
  (\bibinfo{year}{1999}): \emph{\bibinfo{title}{Model checking}}.
\newblock \bibinfo{publisher}{MIT Press}, \doi{10.1002/9780470050118.ecse247}.

\bibitemdeclare{inproceedings}{DF11}
\bibitem{DF11}
\bibinfo{author}{Werner \surnamestart Damm\surnameend} \&
  \bibinfo{author}{Bernd \surnamestart Finkbeiner\surnameend}
  (\bibinfo{year}{2011}): \emph{\bibinfo{title}{Does It Pay to Extend the
  Perimeter of a World Model?}}
\newblock In: {\sl \bibinfo{booktitle}{FM 2011: Formal Methods}},
  \bibinfo{publisher}{Springer}, pp. \bibinfo{pages}{12--26},
  \doi{10.1007/978-3-642-21437-0\_4}.

\bibitemdeclare{misc}{Collins}
\bibitem{Collins}
\bibinfo{author}{Collins~English \surnamestart Dictionary\surnameend}
  (\bibinfo{year}{2022}): \emph{\bibinfo{title}{doxastic}}.
\newblock
  \urlprefix\url{www.collinsdictionary.com/de/worterbuch/englisch/doxastic}.

\bibitemdeclare{misc}{CollinsEpi}
\bibitem{CollinsEpi}
\bibinfo{author}{Collins~English \surnamestart Dictionary\surnameend}
  (\bibinfo{year}{2022}): \emph{\bibinfo{title}{epistemics}}.
\newblock
  \urlprefix\url{www.collinsdictionary.com/de/worterbuch/englisch/epistemics}.

\bibitemdeclare{book}{reasoningA}
\bibitem{reasoningA}
\bibinfo{author}{Ronald \surnamestart Fagin\surnameend},
  \bibinfo{author}{Joseph~Y. \surnamestart Halpern\surnameend},
  \bibinfo{author}{Yoram \surnamestart Moses\surnameend} \&
  \bibinfo{author}{Moshe~Y. \surnamestart Vardi\surnameend}
  (\bibinfo{year}{2003}): \emph{\bibinfo{title}{Reasoning About Knowledge}}.
\newblock \bibinfo{publisher}{MIT Press}, \doi{10.7551/mitpress/5803.001.0001}.

\bibitemdeclare{article}{EpistemicsLogics}
\bibitem{EpistemicsLogics}
\bibinfo{author}{Paolo \surnamestart Galeazzi\surnameend} \&
  \bibinfo{author}{Emiliano \surnamestart Lorini\surnameend}
  (\bibinfo{year}{2016}): \emph{\bibinfo{title}{Epistemic logic meets epistemic
  game theory: a comparison between multi-agent Kripke models and type
  spaces}}.
\newblock {\sl \bibinfo{journal}{Synthese}}
  \bibinfo{volume}{193}(\bibinfo{number}{7}), pp. \bibinfo{pages}{2097--2127},
  \doi{10.1007/s11229-015-0834-x}.

\bibitemdeclare{inbook}{BelRevIntro}
\bibitem{BelRevIntro}
\bibinfo{author}{Peter \surnamestart Gärdenfors\surnameend}
  (\bibinfo{year}{1992}): \emph{\bibinfo{title}{Belief Revision}}, chapter
  \bibinfo{chapter}{Belief Revision: An Introduction}, pp.
  \bibinfo{pages}{1--28}.
\newblock \bibinfo{publisher}{Cambridge University Press},
  \doi{10.1017/CBO9780511526664.001}.

\bibitemdeclare{article}{Autonomics}
\bibitem{Autonomics}
\bibinfo{author}{David \surnamestart Harel\surnameend}, \bibinfo{author}{Assaf
  \surnamestart Marron\surnameend} \& \bibinfo{author}{Joseph \surnamestart
  Sifakis\surnameend} (\bibinfo{year}{2020}): \emph{\bibinfo{title}{Autonomics:
  In search of a foundation for next-generation autonomous systems}}.
\newblock {\sl \bibinfo{journal}{Proceedings of the National Academy of
  Sciences}} \bibinfo{volume}{117}(\bibinfo{number}{30}), pp.
  \bibinfo{pages}{17491--17498}, \doi{10.1073/pnas.2003162117}.

\bibitemdeclare{inproceedings}{ATEL}
\bibitem{ATEL}
\bibinfo{author}{Wiebe \surnamestart van~der Hoek\surnameend} \&
  \bibinfo{author}{Michael \surnamestart Wooldridge\surnameend}
  (\bibinfo{year}{2002}): \emph{\bibinfo{title}{Tractable Multiagent Planning
  for Epistemic Goals}}.
\newblock In: {\sl \bibinfo{booktitle}{Proc. of the 1st Int. Joint Conference
  on Autonomous Agents and Multiagent Systems: Part 3}}, \bibinfo{series}{AAMAS
  '02}, \bibinfo{publisher}{Association for Computing Machinery}, p.
  \bibinfo{pages}{1167–1174}, \doi{10.1145/545056.545095}.

\bibitemdeclare{book}{IntelAdSys}
\bibitem{IntelAdSys}
\bibinfo{author}{Ming \surnamestart Hou\surnameend}, \bibinfo{author}{Simon
  \surnamestart Banbury\surnameend} \& \bibinfo{author}{Catherine \surnamestart
  Burns\surnameend} (\bibinfo{year}{2014}): \emph{\bibinfo{title}{Intelligent
  Adaptive Systems: An Interaction-Centered Design Perspective}}.
\newblock \bibinfo{publisher}{CRC Press}, \doi{10.1201/b17742}.

\bibitemdeclare{misc}{Uber18}
\bibitem{Uber18}
\bibinfo{author}{Timothy~B. \surnamestart Lee\surnameend}:
  \emph{\bibinfo{title}{Self-driving cars -- How terrible software design
  decisions led to Uber's deadly 2018 crash}}.
\newblock
  \urlprefix\url{arstechnica.com/cars/2019/11/how-terrible-software-design-decisions-led-to-ubers-deadly-2018-crash}.

\bibitemdeclare{article}{DELStrat}
\bibitem{DELStrat}
\bibinfo{author}{Bastien \surnamestart Maubert\surnameend},
  \bibinfo{author}{Aniello \surnamestart Murano\surnameend},
  \bibinfo{author}{Sophie \surnamestart Pinchinat\surnameend},
  \bibinfo{author}{Fran{\c{c}}ois \surnamestart Schwarzentruber\surnameend} \&
  \bibinfo{author}{Silvia \surnamestart Stranieri\surnameend}
  (\bibinfo{year}{2020}): \emph{\bibinfo{title}{Dynamic Epistemic Logic Games
  with Epistemic Temporal Goals}}.
\newblock {\sl \bibinfo{journal}{CoRR}} \bibinfo{volume}{abs/2001.07141},
  \doi{10.3233/FAIA200088}.

\bibitemdeclare{incollection}{DoxasticMeyer2003}
\bibitem{DoxasticMeyer2003}
\bibinfo{author}{John-Jules~Ch. \surnamestart Meyer\surnameend}
  (\bibinfo{year}{2003}): \emph{\bibinfo{title}{Modal Epistemic and Doxastic
  Logic}}.
\newblock In: {\sl \bibinfo{booktitle}{Handbook of Philosophical Logic}},
  \bibinfo{publisher}{Springer}, pp. \bibinfo{pages}{1--38},
  \doi{10.1007/978-94-017-4524-6\_1}.

\bibitemdeclare{article}{accidents2020}
\bibitem{accidents2020}
\bibinfo{author}{Dorde \surnamestart Petrovic\surnameend},
  \bibinfo{author}{Radomir \surnamestart Mijailovic\surnameend} \&
  \bibinfo{author}{Dalibor \surnamestart Pesic\surnameend}
  (\bibinfo{year}{2020}): \emph{\bibinfo{title}{Traffic Accidents with
  Autonomous Vehicles: Type of Collisions, Manoeuvres and Errors of
  Conventional Vehicles’ Drivers}}.
\newblock {\sl \bibinfo{journal}{Transport Infrastructure and systems in a
  changing world. Towards a more sustainable, reliable and smarter mobility.
  TIS Roma 2019 Conference Proceedings}} \bibinfo{volume}{45}, pp.
  \bibinfo{pages}{161--168}, \doi{10.1016/j.trpro.2020.03.003}.

\bibitemdeclare{article}{LTLSynth}
\bibitem{LTLSynth}
\bibinfo{author}{A.~\surnamestart Pnueli\surnameend} \& \bibinfo{author}{Roni
  \surnamestart Rosner\surnameend} (\bibinfo{year}{1989}):
  \emph{\bibinfo{title}{On the synthesis of a reactive module}}.
\newblock {\sl \bibinfo{journal}{Automata Languages and Programming}}
  \bibinfo{volume}{372}, pp. \bibinfo{pages}{179--190},
  \doi{10.1145/75277.75293}.

\bibitemdeclare{article}{BDI}
\bibitem{BDI}
\bibinfo{author}{Anand~S. \surnamestart Rao\surnameend} \&
  \bibinfo{author}{Michael~P. \surnamestart Georgeff\surnameend}
  (\bibinfo{year}{1998}): \emph{\bibinfo{title}{{Decision Procedures for BDI
  Logics}}}.
\newblock {\sl \bibinfo{journal}{Journal of Logic and Computation}}
  \bibinfo{volume}{8}(\bibinfo{number}{3}), pp. \bibinfo{pages}{293--343},
  \doi{10.1093/logcom/8.3.293}.

\bibitemdeclare{inproceedings}{HCPS}
\bibitem{HCPS}
\bibinfo{author}{David \surnamestart Romero\surnameend}, \bibinfo{author}{Peter
  \surnamestart Bernus\surnameend}, \bibinfo{author}{Ovidiu \surnamestart
  Noran\surnameend}, \bibinfo{author}{Johan \surnamestart Stahre\surnameend} \&
  \bibinfo{author}{{\AA}sa \surnamestart Fast-Berglund\surnameend}
  (\bibinfo{year}{2016}): \emph{\bibinfo{title}{The Operator 4.0: Human
  Cyber-Physical Systems {\&} Adaptive Automation Towards Human-Automation
  Symbiosis Work Systems}}.
\newblock In: {\sl \bibinfo{booktitle}{Advances in Production Management
  Systems. Initiatives for a Sustainable World}},
  \bibinfo{publisher}{Springer}, pp. \bibinfo{pages}{677--686},
  \doi{10.1007/978-3-319-51133-7\_80}.

\bibitemdeclare{misc}{Sheffield}
\bibitem{Sheffield}
\bibinfo{author}{The~University \surnamestart of~Sheffield\surnameend}
  (\bibinfo{year}{2021}): \emph{\bibinfo{title}{{Epistemology}}}.
\newblock
  \urlprefix\url{www.sheffield.ac.uk/philosophy/research/themes/epistemology}.

\bibitemdeclare{article}{motion}
\bibitem{motion}
\bibinfo{author}{Mingyu \surnamestart Wang\surnameend}, \bibinfo{author}{Zijian
  \surnamestart Wang\surnameend}, \bibinfo{author}{John \surnamestart
  Talbot\surnameend}, \bibinfo{author}{J.~Christian \surnamestart
  Gerdes\surnameend} \& \bibinfo{author}{Mac \surnamestart Schwager\surnameend}
  (\bibinfo{year}{2021}): \emph{\bibinfo{title}{Game-Theoretic Planning for
  Self-Driving Cars in Multivehicle Competitive Scenarios}}.
\newblock {\sl \bibinfo{journal}{IEEE Transactions on Robotics}}
  \bibinfo{volume}{37}(\bibinfo{number}{4}), pp. \bibinfo{pages}{1313--1325},
  \doi{10.1109/TRO.2020.3047521}.

\bibitemdeclare{inproceedings}{Autonomous}
\bibitem{Autonomous}
\bibinfo{author}{Yingxu \surnamestart Wang\surnameend},
  \bibinfo{author}{Konstantinos~N. \surnamestart Plataniotis\surnameend},
  \bibinfo{author}{Sam \surnamestart Kwong\surnameend}, \bibinfo{author}{Henry
  \surnamestart Leung\surnameend}, \bibinfo{author}{Svetlana \surnamestart
  Yanushkevich\surnameend}, \bibinfo{author}{Fakhri \surnamestart
  Karray\surnameend}, \bibinfo{author}{Ming \surnamestart Hou\surnameend},
  \bibinfo{author}{Newton \surnamestart Howard\surnameend},
  \bibinfo{author}{Rodolfo~A. \surnamestart Fiorini\surnameend},
  \bibinfo{author}{Paolo \surnamestart Soda\surnameend},
  \bibinfo{author}{Edward \surnamestart Tunstel\surnameend},
  \bibinfo{author}{Jianmin \surnamestart Wang\surnameend} \&
  \bibinfo{author}{Shushma \surnamestart Patel\surnameend}
  (\bibinfo{year}{2019}): \emph{\bibinfo{title}{On Autonomous Systems: From
  Reflexive, Imperative and Adaptive Intelligence to Autonomous and Cognitive
  Intelligence}}.
\newblock In: {\sl \bibinfo{booktitle}{2019 IEEE 18th International Conference
  on Cognitive Informatics \& Cognitive Computing}}, pp.
  \bibinfo{pages}{7--12}, \doi{10.1109/ICCICC46617.2019.9146038}.

\bibitemdeclare{article}{AutoWat}
\bibitem{AutoWat}
\bibinfo{author}{David \surnamestart Watson\surnameend} \&
  \bibinfo{author}{David \surnamestart Scheidt\surnameend}  
  (\bibinfo{year}{2005}): \emph{\bibinfo{title}{Autonomous systems}}.
\newblock {\sl \bibinfo{journal}{Johns Hopkins APL Technical Digest (Applied
  Physics Laboratory)}} \bibinfo{volume}{26}, pp. \bibinfo{pages}{368--376}.
\newblock \urlprefix\url{www.jhuapl.edu/Content/techdigest/pdf/V26-N04/26-04-Watson.pdf}

\end{thebibliography}
\end{document}